\documentclass[letterpaper, 10 pt, conference]{ieeeconf}  

\IEEEoverridecommandlockouts                              

\usepackage{amsmath} 
\usepackage{amssymb}  
\usepackage{url}
\usepackage{graphicx}
\usepackage[caption=false,font=footnotesize]{subfig}

\title{\LARGE \bf
Real-Time Nonlinear MPC via Sequential Quadratic Programming with Structure-Exploiting ADMM and Interior-Point Methods for Underactuated Double-Pendulum Swing-Up
}

\author{Nick Karydakis$^{1}$ and Konstantinos Chatzilygeroudis$^{1}$
\thanks{$^{1}$Laboratory of Automation and Robotics (LAR) in the Department of Electrical \& Computer Engineering,
       University of Patras, GR-26504 Patras, Greece,
       {\tt\small costashatz@upatras.gr}}%
}

\begin{document}

\maketitle
\thispagestyle{empty}
\pagestyle{empty}

\begin{abstract}
The 4th ``AI Olympics with RealAIGym'' competition, to be held at IJCAI-ECAI 2026 in Bremen, challenges participants to develop a global control policy for swinging up and stabilizing an underactuated two-link system in its upright position. In contrast to previous editions, participants develop and evaluate their control strategies directly on remotely accessible CloudPendulum hardware, with limited interaction time and without prior knowledge of the system’s model parameters. This paper presents an optimal-control-based approach employing real-time nonlinear model predictive control implemented using sequential quadratic programming. The results demonstrate that the proposed SQP-based MPC controller achieves reliable swing-up and stabilization performance, while maintaining robustness against disturbances.
\end{abstract}

\section{Introduction}
\label{sec:introduction}

The \emph{AI Olympics with RealAIGym} competition series aims to
advance the physical intelligence of robotic systems through
standardized real-world control benchmarks
~\cite{wiebe2022realaigym,wiebe2024athletic,wiebe2026robust}. The fourth edition, held in
conjunction with IJCAI 2026 in Bremen, focuses on global swing-up and
stabilization of underactuated double-pendulum systems using remotely
accessible \emph{CloudPendulum} hardware~\cite{kumar2025swinging,wiebe2023open}.
Participants develop their controllers directly on the physical
platform under limited hardware access and without prior knowledge of
the system parameters.

The competition considers two configurations: the Pendubot, actuated
at the shoulder, and the Acrobot, actuated at the elbow. In both cases,
the objective is to swing up the system and stabilize the unstable
upright equilibrium using a single actuator. The task is challenging
because it requires nonlocal energy-building motions, accurate
stabilization, and robustness to model uncertainty and disturbances.

Although model-free reinforcement learning has achieved competitive
performance in previous editions
~\cite{wiebe2024athletic,wiebe2026robust,faust2024velocity}, this work
investigates a model-based optimal-control approach. We propose a
real-time nonlinear Model Predictive Control (NMPC) method based on
Sequential Quadratic Programming (SQP). At each control step, the
finite-horizon nonlinear optimal control problem is approximated by a
structured Quadratic Program (QP), and the first input of the optimized
sequence is applied in a receding-horizon fashion.

The controller uses a structure-exploiting SQP formulation whose QP
subproblems are solved by a stagewise Alternating Direction Method of
Multipliers (ADMM) solver~\cite{stagewise_admm_carpentier} or by a structure exploiting interior point method, HPIPM~\cite{hpipm}. By exploiting the
temporal sparsity of the optimal control problem, the method is suitable
for real-time execution while retaining explicit treatment of nonlinear
dynamics, underactuation, and actuator constraints. The proposed
controller addresses swing-up and stabilization within a single
optimization-based policy and is evaluated on the remotely accessible
competition hardware.

\section{Problem Formulation}
\label{sec:problem}

\subsection{Nonlinear Optimal Control}

We formulate the swing-up and stabilization task as a discrete-time,
finite-horizon nonlinear optimal control problem~\cite{wensing2024optimisation}. Let
$\boldsymbol{x}_k \in \mathbb{R}^{n}$ and
$\boldsymbol{u}_k \in \mathbb{R}^{m}$ denote the state and control
variables at stage $k$, respectively:

\begin{equation}
\begin{aligned}
\min_{\boldsymbol{x}_k,\boldsymbol{u}_k} \quad
    & \ell_f(\boldsymbol{x}_N)
    + \sum_{k=0}^{N-1}
      \ell(\boldsymbol{x}_k,\boldsymbol{u}_k) \\
\text{s.t.} \quad
    & \boldsymbol{x}_0 = \mathcal{X}_{\text{init}}, \\
    & \boldsymbol{x}_{k+1}
      = \boldsymbol{f}(\boldsymbol{x}_k,\boldsymbol{u}_k),
      \quad k=0,\ldots,N-1, \\
    & \boldsymbol{g}(\boldsymbol{x}_k,\boldsymbol{u}_k)
      \leq \boldsymbol{0},
      \quad k=0,\ldots,N-1, \\
    & \boldsymbol{g}_f(\boldsymbol{x}_N)
      \leq \boldsymbol{0},
\end{aligned}
\label{eq:oc}
\end{equation}

where $\ell$ and $\ell_f$ are the running and terminal costs,
$\boldsymbol{f}$ denotes the discretized system dynamics, $\mathcal{X}_{\text{init}}$ is the initial state, and
$\boldsymbol{g}$ and $\boldsymbol{g}_f$ represent path and terminal
constraints.

The problem is solved using Sequential Quadratic Programming (SQP).
At each SQP iteration, the nonlinear costs and constraints are
approximated locally around a nominal trajectory, yielding a
structured Quadratic Program (QP). The resulting search direction is
used to update the state and control trajectories, and the procedure
is repeated until the convergence criteria are satisfied
~\cite{wright1999numerical,stagewise_admm_carpentier}.

\subsection{Stagewise SQP Subproblem}

At each SQP iteration, the cost is approximated quadratically and the
dynamics and constraints are linearized around the nominal trajectory
$(\bar{\boldsymbol{x}}_k,\bar{\boldsymbol{u}}_k)$. The nominal trajectory simply consists of the current best guess\footnote{In our setting, for the first guess we initialize every variable to zero. In other words, we do not pre-compute a trajectory for the MPC.}, and is updated at each SQP iteration using the solution of the QP. This gives the QP
\begin{align}
\min_{\Delta\boldsymbol{x},\Delta\boldsymbol{u}}
&\sum_{k=0}^{N-1}
\left[
\frac{1}{2}
\begin{bmatrix}
\Delta\boldsymbol{x}_k \\
\Delta\boldsymbol{u}_k
\end{bmatrix}^{\!\top}
\boldsymbol{Q}_k
\begin{bmatrix}
\Delta\boldsymbol{x}_k \\
\Delta\boldsymbol{u}_k
\end{bmatrix}
+
\boldsymbol{q}_k^\top
\begin{bmatrix}
\Delta\boldsymbol{x}_k \\
\Delta\boldsymbol{u}_k
\end{bmatrix}
\right]
\nonumber\\
&\quad+
\frac{1}{2}
\Delta\boldsymbol{x}_N^\top
\boldsymbol{Q}_N
\Delta\boldsymbol{x}_N
+
\boldsymbol{q}_N^\top
\Delta\boldsymbol{x}_N
\end{align}
subject to
\begin{align}
\Delta\boldsymbol{x}_{k+1}
&=
\boldsymbol{A}_k\Delta\boldsymbol{x}_k
+
\boldsymbol{B}_k\Delta\boldsymbol{u}_k
+
\boldsymbol{\gamma}_k,
\\
\boldsymbol{D}_k\Delta\boldsymbol{x}_k
+
\boldsymbol{E}_k\Delta\boldsymbol{u}_k
+
\boldsymbol{g}_k
&\leq \boldsymbol{0},
\\
\boldsymbol{D}_N\Delta\boldsymbol{x}_N
+
\boldsymbol{g}_{f,N}
&\leq \boldsymbol{0},
\end{align}

where $\boldsymbol{A}_k$ and $\boldsymbol{B}_k$ are the dynamics
Jacobians, $\boldsymbol{D}_k$ and $\boldsymbol{E}_k$ are the constraint
Jacobians, and
$\boldsymbol{\gamma}_k
=
\boldsymbol{f}
(\bar{\boldsymbol{x}}_k,\bar{\boldsymbol{u}}_k)
-
\bar{\boldsymbol{x}}_{k+1}$
is the dynamics defect.

The QP is solved using a stagewise ADMM method that exploits the
temporal structure of the optimal control problem~\cite{stagewise_admm_carpentier}. The linear systems arising in the
ADMM iterations are solved using an LQR-based recursion, avoiding the
factorization of a single large sparse KKT system. Our solver also integrates HPIPM~\cite{hpipm} as an alternative structure-exploiting QP solver (based on the interior point method).

\section{Method}
\label{sec:method}

\subsection{Inverse-Dynamics Formulation}
The dynamics of the underactuated two-link system are described by the
manipulator equation
\begin{equation}
\boldsymbol{M}(\boldsymbol{q})\dot{\boldsymbol{v}}
+
\boldsymbol{C}(\boldsymbol{q},\boldsymbol{v})
+
\boldsymbol{g}(\boldsymbol{q})
+
\boldsymbol{d}(\boldsymbol{v})
=
\boldsymbol{S}\boldsymbol{\tau},
\label{eq:manipulator}
\end{equation}
where $\boldsymbol{q}\in\mathbb{R}^{2}$ is the joint configuration,
$\boldsymbol{v}\in\mathbb{R}^{2}$ is the joint velocity,
$\boldsymbol{M}(\boldsymbol{q})$ is the inertia matrix,
$\boldsymbol{C}(\boldsymbol{q},\boldsymbol{v})$ contains Coriolis and
centrifugal terms, $\boldsymbol{g}(\boldsymbol{q})$ is the gravity
vector, and $\boldsymbol{d}(\boldsymbol{v})$ is the joint dissipation model. The matrix $\boldsymbol{S}$ selects the actuated joint and
therefore represents either the Pendubot or Acrobot configuration.
The state is defined as
$\boldsymbol{x}_k =
\begin{bmatrix}
\boldsymbol{q}_k^\top &
\boldsymbol{v}_k^\top
\end{bmatrix}^{\!\top}$, while the generalized acceleration is used as the optimization
variable, $\boldsymbol{u}_k = \dot{\boldsymbol{v}}_k$. We use semi-implicit Euler integration,
\begin{align}
\boldsymbol{v}_{k+1}
&=
\boldsymbol{v}_k
+
\Delta t\,\dot{\boldsymbol{v}}_k,
\\
\boldsymbol{q}_{k+1}
&=
\boldsymbol{q}_k
+
\Delta t\,\boldsymbol{v}_{k+1}.
\end{align}
For a given state and generalized acceleration, the required joint
torques are obtained from inverse dynamics as
\begin{equation}
\boldsymbol{\tau}_k
=
\boldsymbol{M}(\boldsymbol{q}_k)\dot{\boldsymbol{v}}_k
+
\boldsymbol{C}(\boldsymbol{q}_k,\boldsymbol{v}_k)
+
\boldsymbol{g}(\boldsymbol{q}_k)
+
\boldsymbol{d}(\boldsymbol{v}_k),
\label{eq:inverse_dynamics}
\end{equation}

which can be efficiently evaluated using the Recursive Newton--Euler Algorithm~\cite{featherstone2008dynamics}.
Because only one joint is actuated, the torque associated with the
passive joint must vanish. This is imposed through the equality
constraint
\begin{equation}
(\boldsymbol{I}-\boldsymbol{S})
\left[
\boldsymbol{M}(\boldsymbol{q}_k)\dot{\boldsymbol{v}}_k
+
\boldsymbol{C}(\boldsymbol{q}_k,\boldsymbol{v}_k)
+
\boldsymbol{g}(\boldsymbol{q}_k)
+
\boldsymbol{d}(\boldsymbol{v}_k)
\right]
=
\boldsymbol{0}.
\label{eq:underactuation_constraint}
\end{equation}

The torque of the actuated joint is constrained by
\begin{equation}
\boldsymbol{\tau}_{\min}
\leq
\boldsymbol{S}
\left[
\boldsymbol{M}(\boldsymbol{q}_k)\dot{\boldsymbol{v}}_k
+
\boldsymbol{C}(\boldsymbol{q}_k,\boldsymbol{v}_k)
+
\boldsymbol{g}(\boldsymbol{q}_k)
+
\boldsymbol{d}(\boldsymbol{v}_k)
\right]
\leq
\boldsymbol{\tau}_{\max}.
\label{eq:torque_limits}
\end{equation}

This formulation allows the actuator limits and the system's
underactuation to be represented explicitly in the SQP subproblem.

\subsection{Real-Time MPC Implementation}

At each control instant, the current state estimate is used to
initialize the finite-horizon optimal control problem. A limited
number of SQP iterations is performed to satisfy the real-time
computational budget. Each SQP iteration constructs a structured QP,
which is solved using either the stagewise ADMM method or the HPIPM QP solver.

The optimized trajectory from the previous control instant is shifted
and used to warm-start both the state-control trajectory and the
associated solver variables. After the SQP iterations are completed,
the torque corresponding to the first optimized acceleration is
computed using~\eqref{eq:inverse_dynamics} and applied to the actuated
joint. The procedure is repeated at the next sampling instant.

The solver parameters used in all MPC experiments are summarized in
Table~\ref{tab:mpc_parameters}. The controller performs four SQP
iterations at each MPC step, with each QP subproblem limited to
$100$ iterations. Warm starting is enabled to reuse the solution
from the previous control cycle.

\begin{table}[t]
    \centering
    \caption{Solver parameters used for the SQP-based MPC controller.}
    \label{tab:mpc_parameters}
    \begin{tabular}{ll}
        \hline
        \textbf{Parameter} & \textbf{Value} \\
        \hline
        SQP tolerance (ADMM) & $1\times10^{-1}$ \\
        SQP tolerance (HPIPM) & $1\times10^{-2}$ \\
        ADMM absolute tolerance & $1\times10^{-2}$ \\
        ADMM relative tolerance & $1\times10^{-2}$ \\
        ADMM (dual vars) warm start & Enabled \\
        HPIPM tolerances & $1\times10^{-3}$ \\
        Maximum SQP iterations per MPC step (ADMM) & $4$ \\
        Maximum SQP iterations per MPC step (HPIPM) & $6$ \\
        Maximum QP iterations per SQP iteration & $100$ \\
        \hline
    \end{tabular}
    \vspace{-2em}
\end{table}

\subsection{Underactuated Double-Pendulum Dynamics}
\label{sec:double_pendulum_dynamics}




The Acrobot and Pendubot are modeled as underactuated planar
two-link manipulators with identical mechanical parameters. They
differ only in the location of the actuator. The generalized
coordinates are defined as
$\boldsymbol{q}
    =
    \begin{bmatrix}
        q_1 & q_2
    \end{bmatrix}^{\top}$,
where $q_1$ is the shoulder angle and $q_2$ is the relative elbow
angle. The downward hanging configuration corresponds to
$\boldsymbol{q}=\boldsymbol{0}$.
We use the following model parameters:
\begin{align}
    m_1 &= 0.10548~\mathrm{kg}, &
    m_2 = 0.0762~\mathrm{kg}, \\
    l_1 &= 0.05~\mathrm{m}, \\
    r_1 &= 0.05~\mathrm{m}, &
    r_2 = 0.0367004~\mathrm{m}, \\
    I_1 &= 4.616622\times10^{-4}~\mathrm{kg\,m^2},\\
    I_2 &= 2.370240\times10^{-4}~\mathrm{kg\,m^2}.
\end{align}
where $l_1$ is the distance between the two joints,
$r_i$ is the distance from joint $i$ to the center of mass of link
$i$, and $I_i$ is the link inertia about the joint rotation axis.
We define
\begin{align}
    a &=
    I_1 + I_2 + m_1 r_1^2
    + m_2\left(l_1^2+r_2^2\right), \\
    b &= m_2 l_1 r_2, \\
    d &= I_2+m_2r_2^2.
\end{align}

The inertia matrix is
\begin{equation}
\boldsymbol{M}(\boldsymbol{q})
=
\begin{bmatrix}
a+2b\cos q_2
&
d+b\cos q_2
\\
d+b\cos q_2
&
d
\end{bmatrix}.
\label{eq:mass_matrix}
\end{equation}

The Coriolis and centrifugal vector is
\begin{equation}
\boldsymbol{C}(\boldsymbol{q},\dot{\boldsymbol{q}})
=
\begin{bmatrix}
-b\sin q_2
\left(
2\dot q_1\dot q_2+\dot q_2^2
\right)
\\[1mm]
b\sin q_2\,\dot q_1^2
\end{bmatrix}.
\label{eq:coriolis_vector}
\end{equation}

Taking the downward configuration as the zero-angle configuration,
the gravity vector is
\begin{equation}
\boldsymbol{g}(\boldsymbol{q})
=
\begin{bmatrix}
\left(m_1r_1+m_2l_1\right)g\sin q_1
+
m_2r_2g\sin(q_1+q_2)
\\[1mm]
m_2r_2g\sin(q_1+q_2)
\end{bmatrix}.
\label{eq:gravity_vector}
\end{equation}

The joint dissipation model is
\begin{equation}
\boldsymbol{d}(\dot{\boldsymbol{q}})
=
\begin{bmatrix}
b_1\dot q_1 + f_1\,\operatorname{sgn}(\dot q_1)
\\
b_2\dot q_2 + f_2\,\operatorname{sgn}(\dot q_2)
\end{bmatrix},
\label{eq:joint_friction}
\end{equation}
with
\begin{align*}
    b_1 &= 7.6341\times 10^{-12}~\mathrm{N\,m\,s/rad},\\
    f_1 &= 3.05\times 10^{-3}~\mathrm{N\,m}, \\
    b_2 &= 5.1065\times 10^{-4}~\mathrm{N\,m\,s/rad},\\
    f_2 &= 7.777\times 10^{-4}~\mathrm{N\,m}.
\end{align*}

All parameter values have been determined by performing system identification on the pendulum hardware.

\subsubsection{Pendubot}

For the Pendubot, the shoulder joint is actuated and the elbow joint
is passive. Hence, the dynamics become
    $\boldsymbol{M}(\boldsymbol{q})\ddot{\boldsymbol{q}}
    +
    \boldsymbol{C}(\boldsymbol{q},\dot{\boldsymbol{q}})
    +
    \boldsymbol{g}(\boldsymbol{q})
    +
    \boldsymbol{d}(\dot{\boldsymbol{q}})
    =
    \begin{bmatrix}
        \tau \\ 0
    \end{bmatrix}$.
\begin{figure}[!htb]
    \centering
    \includegraphics[width=\linewidth]
    {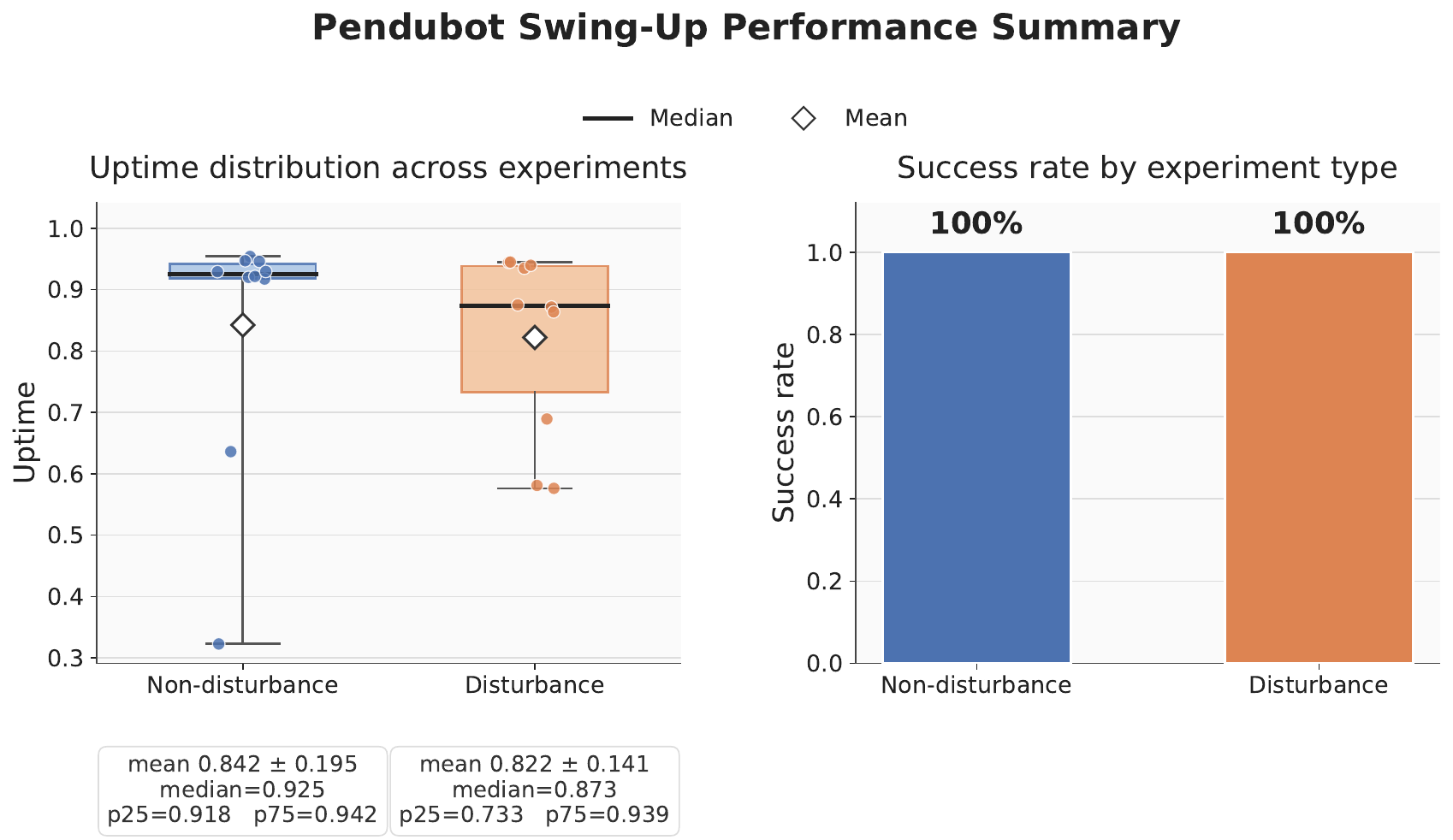}
    \vspace{-2em}
    \caption{Pendubot performance over $60~\mathrm{s}$ episodes (we run 10 independent episodes per scenario). Median
    uptime is $55.5~\mathrm{s}$ ($0.925$) without disturbances and
    $52.38~\mathrm{s}$ ($0.873$) with torque disturbances. Both scenarios achieve a
    $100\%$ swing-up success rate.}
    \label{fig:performance_summary}
    \vspace{-1em}
\end{figure}
\begin{figure}[!htb]
    \centering
    \includegraphics[width=\linewidth]
    {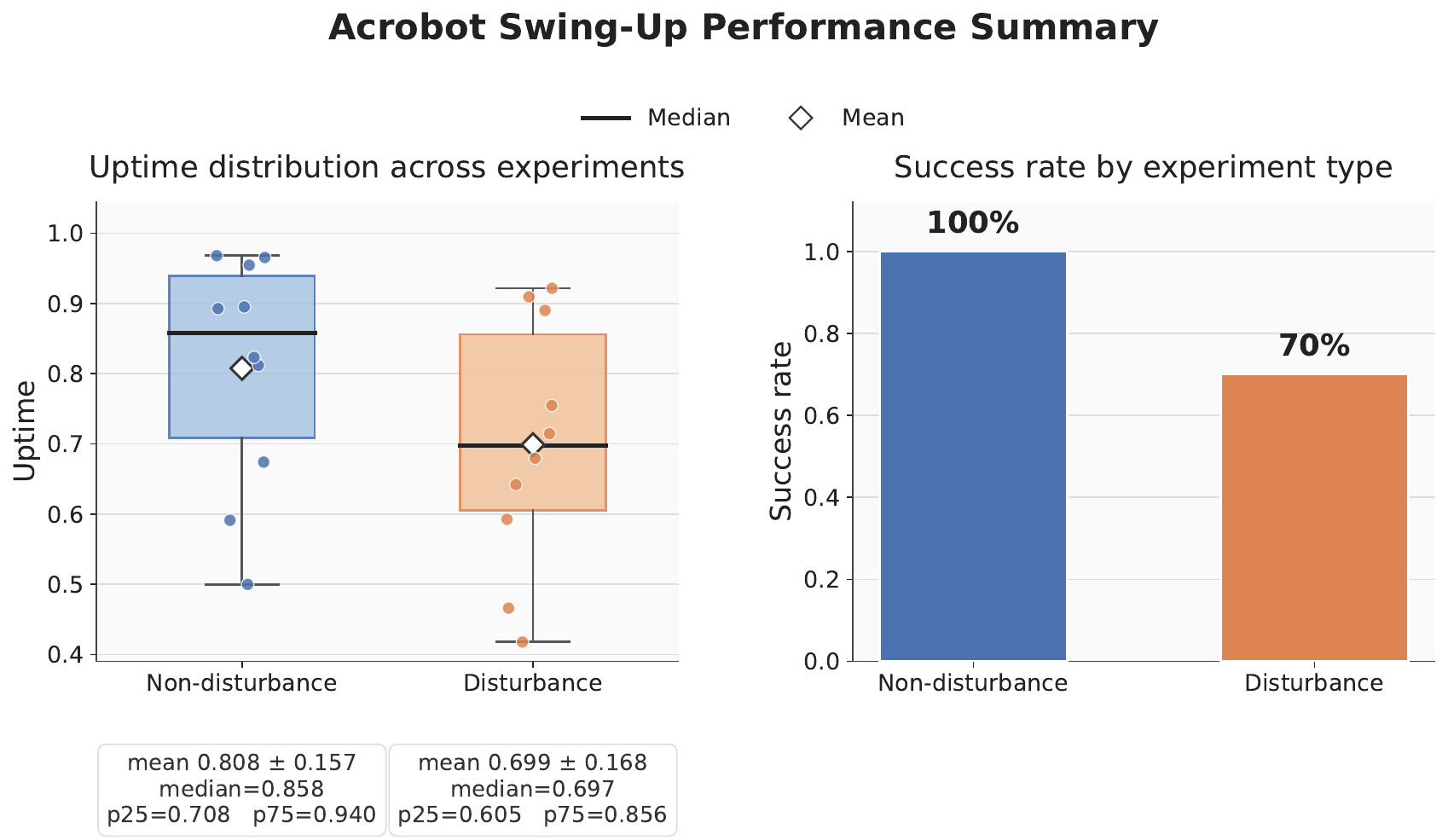}
    \vspace{-2em}
    \caption{Acrobot performance over $60~\mathrm{s}$ episodes (we run 10 independent episodes per scenario). Median
    uptime is $51.48~\mathrm{s}$ ($0.858$) without disturbances and
    $41.82~\mathrm{s}$ ($0.697$) with torque disturbances. We achieve a
    $100\%$ and $70\%$ swing-up success rate respectively.}
    \label{fig:performance_summary_acrobot}
    \vspace{-2em}
\end{figure}

The torque and velocity limits are\footnote{We use limits lower than the actual maximum to make sure that we do not get velocity or torque violations often.}: $|\tau| \leq 0.1~\mathrm{N\,m}, |\dot q_1|,|\dot q_2| \leq 20~\mathrm{rad/s}$.
\subsubsection{Acrobot}
For the Acrobot, the shoulder joint is passive and the elbow joint is
actuated. Therefore, the dynamics become
    $\boldsymbol{M}(\boldsymbol{q})\ddot{\boldsymbol{q}}
    +
    \boldsymbol{C}(\boldsymbol{q},\dot{\boldsymbol{q}})
    +
    \boldsymbol{g}(\boldsymbol{q})
    +
    \boldsymbol{d}(\dot{\boldsymbol{q}})
    =
    \begin{bmatrix}
        0 \\ \tau
    \end{bmatrix}$.
The actuator and velocity limits are: $|\tau| \leq 0.08~\mathrm{N\,m}, |\dot q_1|\leq 30~\mathrm{rad/s},|\dot q_2| \leq 25~\mathrm{rad/s}$. To improve robustness on the real system, we allow a small control input at the nominally passive joint, with its torque limited to $0.005\mathrm{N\,m}$. This effectively provides a small amount of \emph{friction compensation}.
\subsection{MPC Cost Functions}
For both the Pendubot and Acrobot, the model predictive controller penalizes
joint velocities, control effort, and task-space tracking errors\footnote{$\ell_k$ denotes the running cost and $\ell_f$ denotes the terminal cost.}.
The weighted squared norm is:
\begin{align}
\left\|
\boldsymbol{x}
\right\|_{\boldsymbol{W}}^2
=
\boldsymbol{x}^{\top}
\boldsymbol{W}
\boldsymbol{x}.
\end{align}
\subsubsection{Pendubot}
For the Pendubot, the running cost penalizes the joint velocities, control
effort, and tip-position error,
\begin{align}
\ell_k^{\mathrm{P}}
=\;
w_v
\left\|
\boldsymbol{v}_k
\right\|_2^2
+
w_u
\left\|
\boldsymbol{u}_k
\right\|_2^2
+
w_p
\left\|
\boldsymbol{p}_k
-
\boldsymbol{p}_{\mathrm{ref}}
\right\|_2^2,
\label{eq:pendubot_running_cost}
\end{align}
where $\boldsymbol{v}_k$ denotes the joint velocity vector,
$\boldsymbol{u}_k$ denotes the control input, and
$\boldsymbol{p}_k$ is the Cartesian position of the tip.
At the terminal stage, an additional configuration tracking term is included,
such that
\begin{align}
\ell_f^{\mathrm{P}}
&=\;
w_v
\left\|
\boldsymbol{v}_N
\right\|_2^2
+
w_p
\left\|
\boldsymbol{p}_N
-
\boldsymbol{p}_{\mathrm{ref}}
\right\|_2^2
\nonumber\\
&+
\left\|
\boldsymbol{q}_N
-
\boldsymbol{q}_{\mathrm{ref}}
\right\|_{\boldsymbol{W}_q}^2.
\label{eq:pendubot_terminal_cost}
\end{align}
The Pendubot cost weights are selected as
\begin{align}
w_v &= 3\times 10^{-2}, &
w_u &= 2\times 10^{-3}, &
w_p &= 4.2.
\end{align}

The desired tip position is
$\boldsymbol{p}_{\mathrm{ref}}
=
\begin{bmatrix}
0 &
0 &
0.2
\end{bmatrix}^{\top}$,
and the desired joint configuration is
$\boldsymbol{q}_{\mathrm{ref}}
=
\begin{bmatrix}
\pi &
0
\end{bmatrix}^{\top}$.
%
The configuration weighting matrix is
$\boldsymbol{W}_q
=
\begin{bmatrix}
0 & 0 \\
0 & 2
\end{bmatrix}$.
%
\subsubsection{Acrobot}
For the Acrobot, the running cost penalizes the joint velocities, control
effort, and the Cartesian position errors of both the elbow and the tip,
\begin{align}
\ell_k^{\mathrm{A}}
=&\;
\left\|
\boldsymbol{v}_k
\right\|_{\boldsymbol{W}_v^{\mathrm{A}}}^2
+
\left\|
\boldsymbol{u}_k
\right\|_{\boldsymbol{W}_u^{\mathrm{A}}}^2
\nonumber\\
&+
w_e
\left\|
\boldsymbol{p}_{e,k}
-
\boldsymbol{p}_{e,\mathrm{ref}}
\right\|_2^2
\nonumber\\
&+
w_t
\left\|
\boldsymbol{p}_{t,k}
-
\boldsymbol{p}_{t,\mathrm{ref}}
\right\|_2^2,
\label{eq:acrobot_running_cost}
\end{align}
where $\boldsymbol{p}_{e,k}$ and $\boldsymbol{p}_{t,k}$ denote the Cartesian
positions of the elbow and tip, respectively.
The joint-velocity and control weighting matrices are selected as
\begin{align}
\boldsymbol{W}_v^{\mathrm{A}}
&=
\begin{bmatrix}
0.6 & 0 \\
0 & 0.24
\end{bmatrix},
&
\boldsymbol{W}_u^{\mathrm{A}}
&=
\begin{bmatrix}
0 & 0 \\
0 & 3\times 10^{-2}
\end{bmatrix}.
\end{align}

The elbow- and tip-position tracking weights are
\begin{align}
w_e &= 200, &
w_t &= 120.
\end{align}

The corresponding Cartesian reference positions are
\begin{align}
\boldsymbol{p}_{e,\mathrm{ref}}
&=
\begin{bmatrix}
0 &
0 &
0.1
\end{bmatrix}^{\top},
&
\boldsymbol{p}_{t,\mathrm{ref}}
&=
\begin{bmatrix}
0 &
0 &
0.2
\end{bmatrix}^{\top}.
\end{align}
At the terminal stage, the control-effort term is omitted, while the velocity,
elbow-position, and tip-position penalties are retained. The Acrobot terminal
cost is therefore
\begin{align}
\ell_f^{\mathrm{A}}
=&\;
\left\|
\boldsymbol{v}_N
\right\|_{\boldsymbol{W}_v^{\mathrm{A}}}^2
+
w_e
\left\|
\boldsymbol{p}_{e,N}
-
\boldsymbol{p}_{e,\mathrm{ref}}
\right\|_2^2
\nonumber\\
&+
w_t
\left\|
\boldsymbol{p}_{t,N}
-
\boldsymbol{p}_{t,\mathrm{ref}}
\right\|_2^2.
\label{eq:acrobot_terminal_cost}
\end{align}
\section{Experimental Results}
\label{sec:results}
To solve the MPC problem in real time, we utilize our custom SQP solver implementation, \textit{SSQP}\footnote{We plan to release the source code in the immediate future.}. We tested the controller in both the Pendubot and the Acrobot systems. In the final configuration, we use the ADMM solver for the Pendubot and HPIPM for the Acrobot, as these choices yielded the best performance after tuning.
%
%
All experiments consisted of $60~\mathrm{s}$ episodes initialized
from the downward hanging configuration:
$
    \boldsymbol{q}_0 =
    \begin{bmatrix}
        0 & 0
    \end{bmatrix}^{\top},
    \dot{\boldsymbol{q}}_0 =
    \begin{bmatrix}
        0 & 0
    \end{bmatrix}^{\top}$.

Two experimental scenarios were considered. In the first scenario,
the controller performed the swing-up and stabilization task without
external disturbances. In the second scenario, external torque
disturbances were applied after the system had reached the upright
configuration, testing the controller's ability to reject the
disturbances and recover upright stabilization. The disturbances consisted of the application of a random torque to the actuated joint for a period of $\mathrm{0.1}~\mathrm{s}$. For the Pendubot system we apply the disturbance at experiment times $t = \mathrm{20}~\mathrm{s}$ and $t = \mathrm{40}~\mathrm{s}$, while for the Acrobot system we only apply the second one.

A trial was considered successful when the controller swung the
Pendubot up from the downward configuration and maintained it around
the upright equilibrium for at least $5~\mathrm{s}$. For the disturbance experiments, a trial was considered successful
when the controller recovered from the applied torque disturbances and
subsequently maintained the Pendulum around the upright equilibrium
for at least $5~\mathrm{s}$. We performed 10 independent episodes/runs for each scenario (aka, 20 in total per system), and we report median and 25\%, 75\% percentiles.
In addition to the success rate, we evaluate the fraction of each
episode during which the system remains within the upright
stabilization region. We refer to this metric as the
\emph{uptime}. Let $\mathcal{X}_{\mathrm{up}}$ denote the set of
states considered upright and stabilized. For an episode of duration
$T_{\mathrm{ep}}=60~\mathrm{s}$, the uptime is defined as
\begin{equation}
    U
    =
    \frac{1}{N_{\mathrm{ep}}}
    \sum_{k=0}^{N_{\mathrm{ep}}-1}
    \mathbb{I}
    \left[
        \boldsymbol{x}_k\in\mathcal{X}_{\mathrm{up}}
    \right].
    \label{eq:uptime_discrete}
\end{equation}
where $\mathbb{I}[\cdot]$ denotes the indicator function.

\subsection{Pendubot Results}\label{sec:pendubot_results}
In both scenarios, our controller achieved 100\% success rate (Fig.~\ref{fig:performance_summary}). For the experiments without disturbances, the controller achieved an
average uptime (median value) of $\mathrm{92.5}\%$, corresponding to an average of
$\mathrm{55.5}~\mathrm{s}$ in the upright stabilization region
during each $60~\mathrm{s}$ episode. In the experiments with applied
torque disturbances, the average uptime was $\mathrm{87.3}\%$,
corresponding to $\mathrm{52.38}~\mathrm{s}$ per episode. The lower
uptime in the disturbance scenario accounts for the temporary
departure from the upright region and the subsequent recovery motion.

In these experiments, the frequency that the controller is able to operate at is greater than $\mathrm{100}~\mathrm{Hz}$, with a mean of around $\mathrm{400}~\mathrm{Hz}$ on the CloudPendulum platform.
\begin{figure*}[t]
    \centering

    \subfloat[Nominal swing-up and stabilization without external
    disturbances.\label{fig:nominal_swingup}]{
        \includegraphics[width=0.44\textwidth]
        {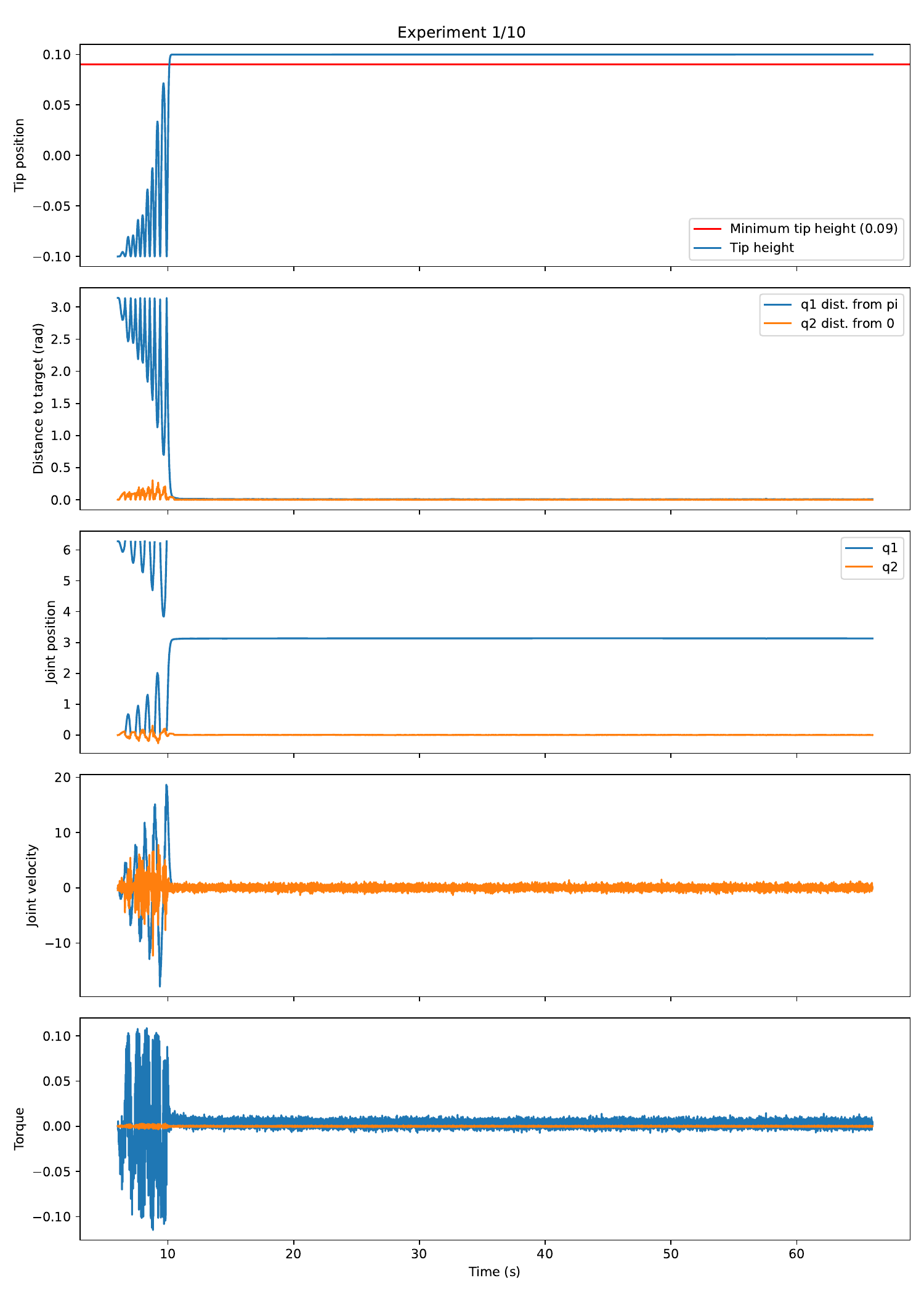}
    }
    \hfill
    \subfloat[Swing-up, stabilization, and recovery after applied
    torque disturbances.\label{fig:disturbed_swingup}]{
        \includegraphics[width=0.44\textwidth]
        {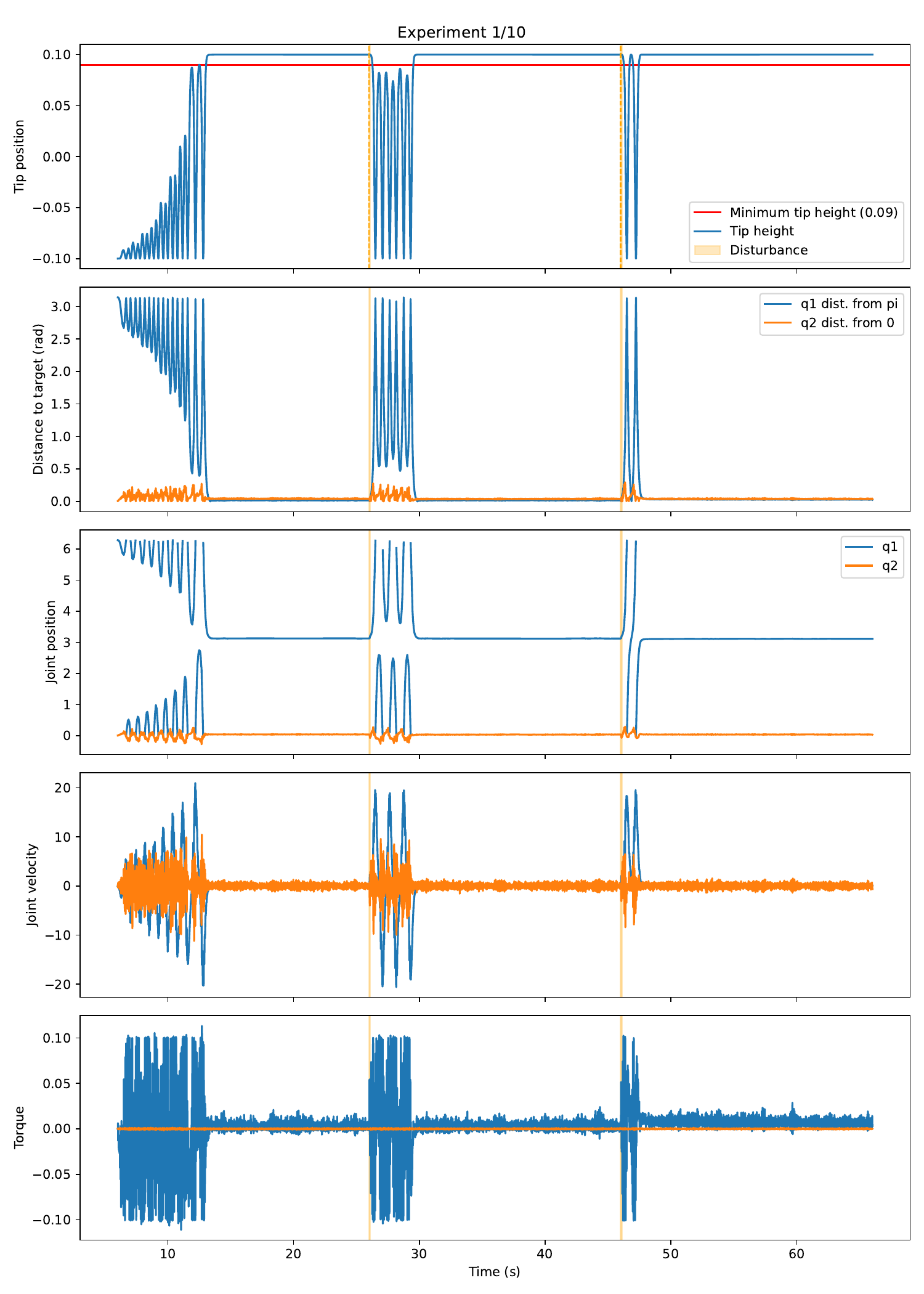}
    }
    \vspace{-0.8em}
    \caption{Representative $60~\mathrm{s}$ Pendubot experiments
    starting from the downward configuration: (a) nominal swing-up
    and stabilization and (b) recovery after applied torque
    disturbances. The controller returns the system to the upright
    equilibrium after each disturbance.}
    \label{fig:representative_results}
    \vspace{-2em}
\end{figure*}
\subsection{Acrobot Results}\label{sec:acrobot_results}
In the Acrobot system, our controller achieved success rates of 100\% and 70\% in the absence and presence of external disturbances, respectively (Fig.~\ref{fig:performance_summary_acrobot}). In the experiments without disturbances, the controller maintained an average uptime of $\mathrm{85.8}\%$ (median value), corresponding to an average of $\mathrm{51.48}~\mathrm{s}$ within the upright stabilization region over each $60~\mathrm{s}$ episode. When torque disturbances were applied, the average uptime decreased to $\mathrm{69.7}\%$, equivalent to $\mathrm{41.82}~\mathrm{s}$ per episode.

In these experiments, the controller operates at frequencies above $\mathrm{100}~\mathrm{Hz}$, with an average frequency of approximately $\mathrm{240}~\mathrm{Hz}$ on the CloudPendulum platform.
\begin{figure*}[t]
    \centering

    \subfloat[Nominal swing-up and stabilization without external
    disturbances.\label{fig:nominal_swingup_acrobot}]{
        \includegraphics[width=0.44\textwidth]
        {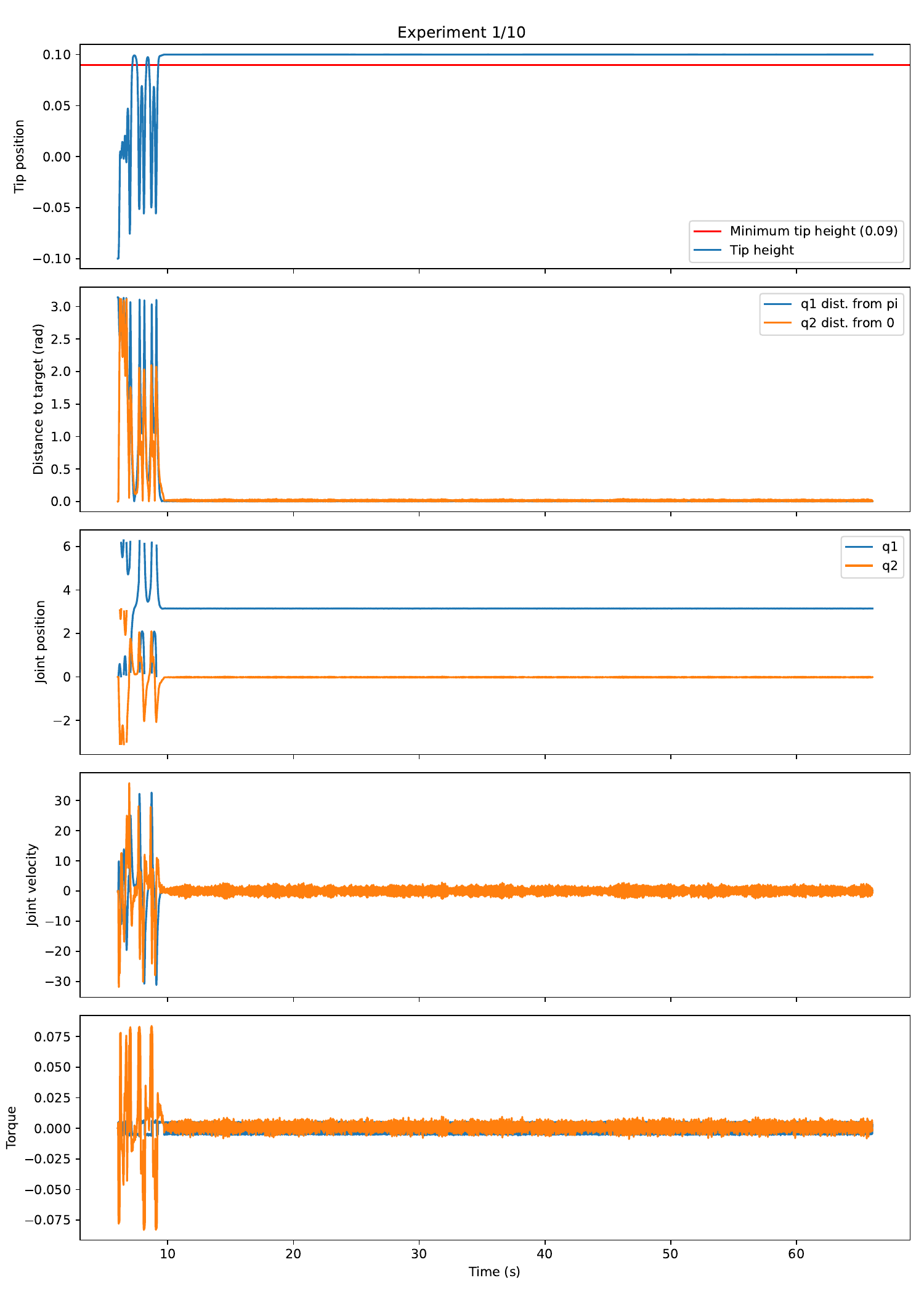}
    }
    \hfill
    \subfloat[Swing-up, stabilization, and recovery after applied
    torque disturbances.\label{fig:disturbed_swingup_acrobot}]{
        \includegraphics[width=0.44\textwidth]
        {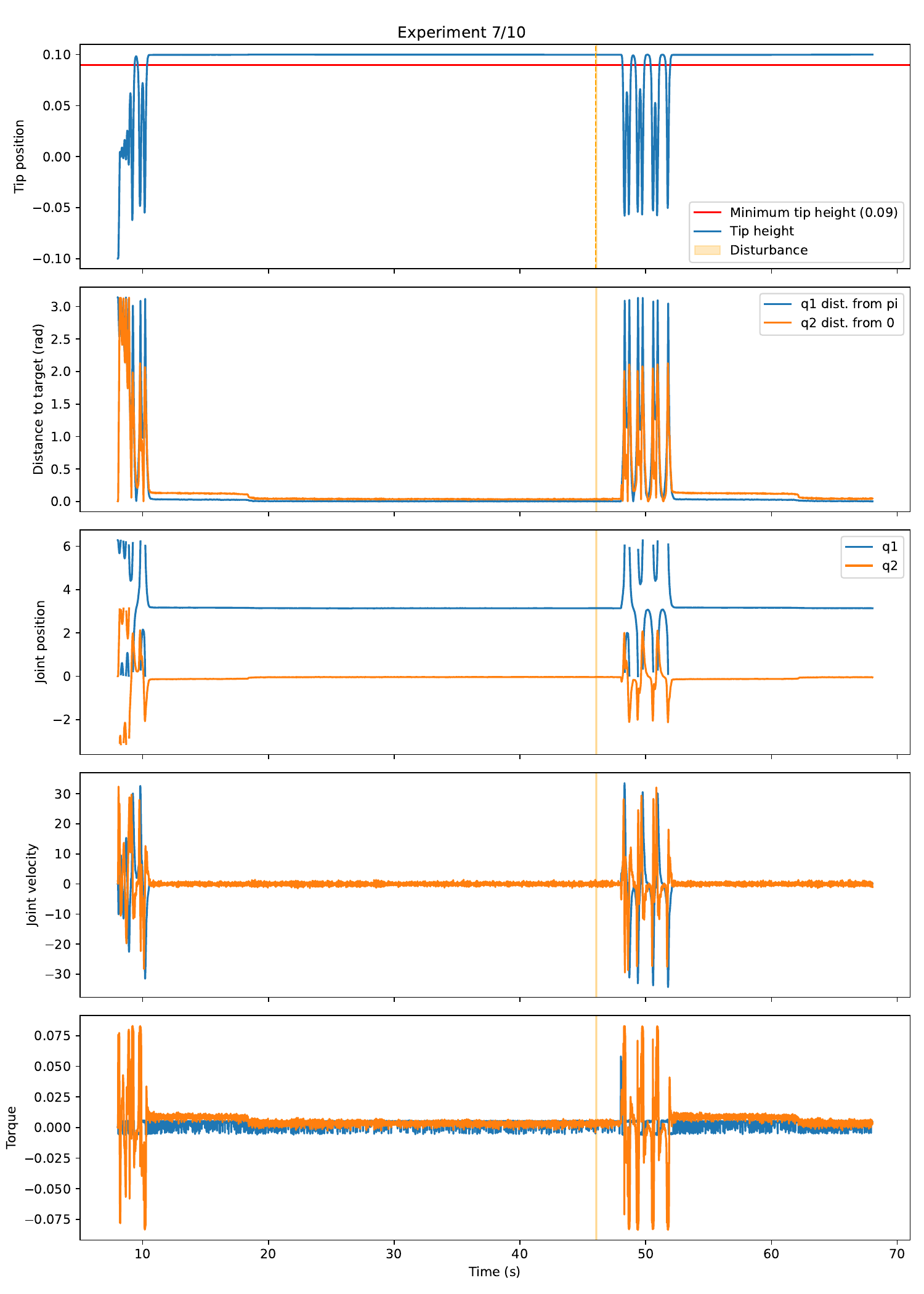}
    }
    \vspace{-0.8em}
    \caption{Representative $60~\mathrm{s}$ Acrobot experiments
    starting from the downward configuration: (a) nominal swing-up
    and stabilization and (b) recovery after applied torque
    disturbances. The controller returns the system to the upright
    equilibrium after each disturbance.}
    \label{fig:representative_results_acrobot}
    \vspace{-2em}
\end{figure*}
\subsection{Verdict}
These results show that the proposed controller can reliably swing up
the Pendubot and Acrobot systems from the downward configuration and maintain the unstable
upright equilibrium over extended runs. The results under torque
disturbances further demonstrate that the receding-horizon controller
can recover from substantial deviations and restore upright
stabilization. See Fig.~\ref{fig:representative_results},~\ref{fig:representative_results_acrobot} for representative trajectories.

The trajectories of all the experiments and the accompanying videos are available at {\footnotesize\url{https://drive.google.com/drive/folders/1ca2ZxxdHlbdJp-Bbk62l5dJV9BryB4rG?usp=sharing}}.
The code for replicating our experiments is available at {\footnotesize\url{https://github.com/upatras-lar/2026_ai_olympics_underactuated_dp}}.

\section{Conclusion and Future Work}
\label{sec:conclusion}

This work presented an inverse-dynamics, SQP-based nonlinear model
predictive controller for the swing-up and stabilization task of the
fourth \emph{AI Olympics with RealAIGym} competition. The controller
was evaluated on the Pendubot and Acrobot systems, with all trials starting
from the downward hanging position. The experimental results show a
swing-up success rate greater than $90\%$ on average, and the controller also demonstrated the ability to
recover from applied torque disturbances and return the system to the
upright position.

The results indicate that the proposed structure-exploiting MPC
approach can generate the nonlocal motions required for swing-up while
also providing feedback stabilization and disturbance rejection.
Furthermore, the controller can execute within the real-time
computational requirements of the remotely accessible CloudPendulum
platform.

Currently, we need \emph{friction compensation} to achieve reliable performance on the Acrobot system. We plan to investigate methods to enable the reliable execution of our MPC on the Acrobot hardware without friction compensation.
%
%
%
Another direction of future work is the integration of online system
identification. The competition model parameters are not known
beforehand and may differ between the development and final evaluation
systems. Online identification will therefore be used to estimate or
adapt relevant dynamic parameters from measured state and control
trajectories. The updated model can then be incorporated into the MPC
prediction model, reducing model mismatch and potentially improving
swing-up reliability, stabilization accuracy, and recovery from
disturbances. Care must be taken since the tuned MPC weights/parameters might need online adaptation.


\bibliographystyle{ieeetr}
\bibliography{references}

\end{document}